\documentclass[showpacs,preprintnumbers,amsmath,amssymb,pra,twocolumn]{revtex4}

\usepackage{amsmath}
\usepackage{amssymb}
\usepackage[dvips]{graphicx}
\usepackage[dvips]{color}
\usepackage[english]{babel}

\newcommand{\ket}[1]{\left\vert#1\right\rangle}

\begin{document}

\title{Self-calibrating Tomography for Angular Schmidt Modes in Spontaneous Parametric Down-Conversion}
\author{S.S.Straupe$^1$}\email{straups@yandex.ru} \author{D.P.Ivanov$^1$, A.A.Kalinkin$^1$, I.B.Bobrov$^1$, S.P.Kulik$^1$ and D.Mogilevtsev$^{2,3}$}
\affiliation{$^1$Faculty of Physics, M.V.Lomonosov Moscow State
University, 199001, Moscow, Russia \\
$^2$Centro de Ci\^encias Naturais e Humanas, Universidade Federal
do ABC, Santo Andr\'e,  SP, 09210-170 Brazil;
\\
 $^3$Institute of Physics, Belarus National Academy of Sciences,
F.Skarina Ave. 68, Minsk 220072 Belarus}

\date{\today}

\begin{abstract}
We report an experimental self-calibrating tomography scheme for
entanglement characterization in high-dimensional quantum systems
using Schmidt decomposition techniques. The self-tomography
technique based on maximal likelihood estimation was developed for
characterizing non-ideal measurements in Schmidt basis allowing us
to infer both Schmidt eigenvalues and detecting efficiencies.
\end{abstract}

 \maketitle

\section{Introduction}
\label{introduction}

Building an experimental set-up for performing sophisticated measurements (for example, such as ones required for
performing quantum tomography), one generally needs calibrating
it. For the signals on the single-particle level the task is quite
challenging, especially if one cannot easily use a pre-calibrated
etalon detectors and/or signal sources for the purpose. More than
30 years ago D. N. Klyshko had outlined an
efficient way to solve this problem using the fact that quantum
features of the signal (such as, for example, character of the
photon number distribution) can be used as a precise measurement
tool for a calibration. In particular, D. N. Klyshko had suggested
using the fact of photon pairs creation in the process of
down-conversion for performing an "absolute calibration" of the
detecting scheme \cite{klyshko80}. A detection of one photon of
the pair in one arm of the "absolute calibration" scheme means
that there is the second photon of the pair going through the
other arm. Thus, the ratio of registered counts in both arms gives
an experimentalist efficiency of the detecting set-up installed in
the second arm of the scheme without any pre-calibrated
detector/source. This idea was actively developed and implemented
(see, for example, Refs.
\cite{{sergienko81},{dariano},{genovese},{migdall2007}}).
Recently, this idea had also given rise to more general concept of
"self-calibration" as simultaneous inference of parameters of both
the measurement scheme and the signal
\cite{mogilevtsev2009,mogilevtsev2010,mogilevtsev2012}. Very
recently the first experimental realization of the
self-calibration scheme was presented using polarization-encoded
one- and two-photon states \cite{branczyk2012}. There, the unknown
rotation angle of the measurement basis was recovered together
with the density matrix of the signal.

Here we present the first example of experimental self-calibrating tomography, when a set of parameters describing efficiency of the detecting
scheme is actually inferred together with the parameters of the spatial state of entangled photon pairs generated in the process of spontaneous
parametric down-conversion (SPDC).

Spatial entanglement in SPDC was a subject of intense research during the last decade. Besides
fundamental issues, spatial states of biphoton pairs offer a
platform for high-dimensional quantum states engineering
motivating this interest. One can distinguish two complementary
approaches to spatial qudit engineering with biphotons: one using
"pixel entanglement" and similar schemes
\cite{HowellPRL04,HowellPRL,PaduaPRL,Fedorov}, and another one
based on using high-order coherent (usually Laguerre-Gaussian)
modes
\cite{ZeilingerNature01,PadgettPRL02,MonkenPRA04,VaziriPRL02,VaziriPRL03,TornerPRL,LangfordPRL04,WoerdmanPRL05,PadgettPRL10,vanExterPRA07,vanExterPRL10,BoydPRL10}.
In both approaches achievable dimensionality and collection
efficiency are figures of merit. Dimensionality of effective
Hilbert space is limited by degree of spatial entanglement. In
pixel entanglement schemes, for example, the pixel size should be
made smaller than the coherence radius of the pump in the far
zone, and since the pump is always divergent, even a plain wave,
selected by point-like aperture would be correlated to a whole set
of plain-wave modes in the conjugate beam. The same holds in
general for other possible choices of modes.

It is remarkable that there is a "preferred" basis among the multitude of possible coherent spatial modes, which consists only of pairwise
correlated modes. It is a  set of Schmidt modes. Since it was used for the first time by Law and Eberly \cite{EberlyPRL04}, it has become a common
tool for entanglement analysis of infinite dimensional systems in general, and of spatial states of photons in particular. A direct experimental
attempt to address spatial entanglement of SPDC biphotons in Schmidt basis was made in the recent work of authors \cite{StraupePRA11}. The technique
of projective measurements used in that work suffers from poor quality of spatial mode transformations, resulting in non-ideal measurement scheme.
Here we use self-calibration to account for this non-ideality. We present a self-consistent analysis of the data collected by measuring approximate
Schmidt modes via the set-up similar to Ref.\cite{StraupePRA11}, and demonstrate that self-calibrating tomography is a feasible and practical way to
update both the information about the measurement scheme and Schmidt eigenvalues starting from very general assumptions about them.

The paper is organized as follows: in Section II we briefly describe the main features of SPDC angular spectrum with emphasis on spatial
entanglement and Schmidt decomposition, Section III describes the general concept of the self-calibrating tomography scheme and its particular
realization for inferring Schmidt eigenvalues. Section IV gives the detailed description of our experiments with spatial Schmidt modes. Section V
describes the practical implementation of the self-calibration scheme.

\section{Analyzing spatial entanglement of SPDC biphotons with Schmidt decomposition}
\label{SPDC_state}

Biphotons generated in the SPDC process have continuous frequency
and angular spectrum. Let us consider its structure in some
details. SPDC can be phenomenologically described using the
following effective interaction Hamiltonian \cite{Klyshko}:
\begin{equation}\label{SPDC Hamiltonian}
    H=\int\limits_{V}d^3\vec{r}\chi^{(2)}(\vec{r})E_p^{(-)}(\vec{r})E^{(+)}(\vec{r}))E^{(+)}(\vec{r})+\text{H.c.}
\end{equation}
Here $E_p$ is the classical amplitude of the pump field and $E$ is
the scattered field operator. Considering pump to be
monochromatic, the first order of perturbation theory gives the
following expression for the state of the scattered field:
\begin{equation}\label{SPDC state}
\begin{split}
    \ket{\Psi}=\ket{vac}+\int{d\vec{k_1}d\vec{k_2}\Psi(\vec{k_1},\vec{k_2})\ket{1}_{k_1}\ket{1}_{k_2}},\\
    \Psi(\vec{k_1},\vec{k_2})=\int\limits_V{d^3\vec{r}}\chi^{(2)}(\vec{r})E_p^{(-)}(\vec{r})\exp\left[i\vec{\Delta}\vec{r}\right],
\end{split}
\end{equation}
where $\vec{\Delta}=\vec{k_1}+\vec{k_2}-\vec{k_p}$, $\omega_1+\omega_2=\omega_p$. In the case of collinear phase-matching and under wide crystal approximation one can obtain the following biphoton field amplitude \cite{BurlakovPRA97,MonkenPRA98,MonkenPRA03}:
\begin{equation}\label{Biphoton amplitude}
    \Psi(\vec{k_1},\vec{k_2})=\mathcal{E}_p(\vec{k_{1\perp}}+\vec{k_{2\perp}})\mathcal{F}(\vec{k_{1\perp}}-\vec{k_{2\perp}}),
\end{equation}
where $\mathcal{E}_p(\vec{k_{1\perp}}+\vec{k_{2\perp}})$ stands
for angular spectrum of the pump, and
$\mathcal{F}(\vec{k_{1\perp}}-\vec{k_{2\perp}})$ is a geometrical
factor determined by phase-matching conditions.

Authors of \cite{MonkenPRA98,MonkenPRA03} give the following expression for $\mathcal{F}$:
\begin{equation}\label{Monken}
    \Psi(\vec{k_1},\vec{k_2})=\mathcal{N}\mathcal{E}_p(\vec{k_{1\perp}}+\vec{k_{2\perp}})\mathrm{sinc}\left[\frac{L(\vec{k_{1\perp}}-\vec{k_{2\perp}})^2}{4k_p}\right],
\end{equation}
with $L$ being the crystal length and $\mathcal{N}$ -- a normalization constant. Although this expression is strictly valid only in the case of
small pump divergence \cite{Fedorov,FedorovPRA08}, it appropriately describes the SPDC spectrum in our experiments.

\sloppy{The most developed approach to quantitative analysis of spatial (and frequency) entanglement of SPDC biphoton states is based on using coherent modes decomposition. Biphoton spatial state space is "discretized" by switching from continuous distributions in plane-wave basis of the previous section to discrete distributions in a chosen basis of spatial mode functions $\xi_i(\vec{k}_{1\perp,2\perp})$. For an arbitrary choice of mode functions in the decomposition of spatial state for each of the photons, the biphoton amplitude takes the following form:}
\begin{equation}\label{Mode-decomposition}
    \Psi(\vec{k}_{1\perp},\vec{k}_{2\perp}) = \sum_{i,j=0}^{\infty} {C_{ij}\xi_i(\vec{k}_{1\perp})\xi_j(\vec{k}_{2\perp})}.
\end{equation}

It turns out, that by appropriate choice of the basis mode functions one can transform the expression (\ref{Mode-decomposition}) to a single-sum form
\begin{equation}\label{Schmidt-decomposition}
    \Psi(\vec{k}_{1 \perp},\vec{k}_{2 \perp}) = \sum_{i=0}^{\infty} {\sqrt{\lambda_{i}}\psi_i(\vec{k}_{1\perp})\psi_i(\vec{k}_{2\perp})}
\end{equation}
which is called \emph{Schmidt decomposition}. In this case the basis functions $\psi_i(\vec{k}_{1,\perp})$ should be eigenfunctions of single-photon density matrix $\rho_{1,2}(\vec{k}_{1\perp},\vec{k}_{1\perp})$, and $\lambda_i$ are the corresponding eigenvalues. It means that the appropriate mode functions may be found by solving the following integral equation:
\begin{equation}\label{kernel}
    \int{\rho_{1,2}(\vec{k}_{\perp},\vec{k}_{\perp}')\psi_i(\vec{k}_{\perp}')\mathrm{d}\vec{k}_{\perp}'=\lambda_i\psi_i(\vec{k}_{\perp})}.
\end{equation}
"Average number of Schmidt modes" determined by \emph{Schmidt number}:
\begin{equation}\label{Schmidt-number}
    K = \frac{1}{\sum_{i=0}^{\infty} {\lambda_i^2}}.
\end{equation}
is widely used as an operational measure of entanglement\cite{EberlyPRL04}.

Degree of spatial entanglement of SPDC biphotons described by
wavefunction (\ref{Monken}) was analyzed by Law and Eberly in
\cite{EberlyPRL04}. The pump was assumed to be gaussian
$\mathcal{E}_p=\exp\left[-\frac{|\vec{k_{1\perp}}+\vec{k_{2\perp}}|}{\sigma^2}\right]$.
Authors derived an analytical expression for Schmidt number by
approximating $\mathcal{F}(\vec{k_{1\perp}}-\vec{k_{2\perp}})$
function of (\ref{Monken}) by a gaussian function:
\begin{equation}\label{Gaussian Shmidt number}
    K_g=\frac{1}{4}\left(b\sigma+\frac{1}{b\sigma}\right)^2,
\end{equation}
where $b=\sqrt{\frac{L}{4k_p}}$ is the waist of the gaussian
function modeling
$\mathcal{F}(\vec{k_{1\perp}}-\vec{k_{2\perp}})$. In the following
we will call this procedure "a double gaussian approximation". The
value of $K_g$ is determined by a single parameter $b\sigma$ and
may reach very high values for $b\sigma\gg1$ and $b\sigma\ll1$.
Numerical calculations performed for
\begin{equation}
    \mathcal{F}(\vec{k_{1\perp}}-\vec{k_{2\perp}})=\mathrm{sinc}\left[b(\vec{k_{1\perp}}-\vec{k_{2\perp}})^2\right]
\end{equation}
showed that real value of $K$ is larger than $K_g$ for all values of $b\sigma$.

The choice of mode functions set in (\ref{Mode-decomposition}) is
quite arbitrary and may be determined by convenience in analyzing
a particular physical situation. If the situation corresponds to a
beam propagating in free space, it is natural to chose the
solutions of paraxial wave equation, i.e. Hermite-Gaussian or
Laguerre-Gaussian modes, as a set of mode functions. Moreover it
turns out, that these functions form the Schmidt decomposition for
SPDC with the gaussian pump of moderate divergence.

Let us choose the set of Hermite-Gaussian modes as a basis for decomposition:
\begin{equation}\label{HG-decomposition}
\begin{split}
    &\mathrm{HG}_{nm}(k_x, k_y) \propto \\
    &\mathrm{H}_n\left(\frac{k_x^2}{(\Delta k_x)^2}\right) \mathrm{H}_m\left(\frac{k_y^2}{(\Delta k_y)^2}\right) \exp\left(-\frac{k_x^2 + k_y^2}{2(\Delta k_\bot)^2}\right),
\end{split}
\end{equation}
where $\mathrm{H}_n(x)$ are Hermite polynomials, and $\{k_x,
k_y\}$ are transverse wave-vector components. One can get rid of
two indexes in decomposition (\ref{HG-decomposition}) and
transform it to a form of Schmidt decomposition using
double-gaussian approximation \cite{EberlyPRL04,FedorovJPB09}. For small $\sigma$ we
can make a substitution
$\mathrm{sinc}(\frac{x^2}{\sigma^2})\rightarrow
\exp(-\gamma\frac{x^2}{\sigma^2})$, where $\gamma$ is a
coefficient chosen to make both functions "close" to each other. A
good approximation is provided by choosing a value of
$\gamma=0.86$.\footnote{We have tried several criteria for
choosing $\gamma$, for example equality of FWHM for both
functions, which is used in \cite{FedorovJPB09} and gives
$\gamma=0.249$. However, $\gamma=0.86$ better describes
experimental data, which is not completely understood, but may be
caused by limited angular detection aperture.} The biphoton
wavefunction now takes the following form:
\begin{equation}\label{double-gaussian}
    \Psi(\vec{k_1},\vec{k_2})\propto \exp(-\frac{(\vec{k_{1\perp}}+\vec{k_{2\perp}})^2}{2a^2})\exp\left(-\frac{(\vec{k_{1\perp}}-\vec{k_{2\perp}})^2}{2b^2}\right),
\end{equation}
where $a$ determines the angular bandwidth of the pump, and $b=\sqrt{4k_p/\gamma L}$ -- the phase-matching bandwidth. Since the wavefunction is a product of functions depending only on $k_{1,2x}$ and only on $k_{1,2y}$, it is sufficient to consider the problem in one dimension:
\begin{equation}\label{double-gaussian-1D}
\begin{split}
    &\Psi(k_{1x},k_{2x})= \\
    &\sqrt{\frac{2}{\pi ab}} \exp(-\frac{(k_{1x}+k_{2x})^2}{2a^2})\exp\left(-\frac{(k_{1x}-k_{2x})^2}{2b^2}\right).
\end{split}
\end{equation}
One can show, that solutions of (\ref{kernel}) for such a wave-function have the form \cite{FedorovJPB09}:
\begin{equation}\label{HG-Schmidt-modes}
    \psi_n(k_{1x,2x})=\left(\frac{2}{ab}\right)^{1/4} \phi_n \left(\sqrt{\frac{2}{ab}} k_{1x,2x}\right),
\end{equation}
where $\phi_n(x)= (2^n n!\sqrt{\pi})^{-1/2}e^{-x^2/2}\mathrm{H}_n(x)$. For corresponding eigenvalues and Schmidt number we obtain:
\begin{equation}\label{HG-Schmidt-eigenvalues}
    \lambda_n = 4ab\frac{(a-b)^{2n}}{(a+b)^{2(n+1)}}, \quad K_{x} = \frac{a^2+b^2}{2ab}.
\end{equation}
So we have the following form of Schmidt decomposition for SPDC biphoton state under the double-gaussian approximation:
\begin{equation}\label{HG-Schmidt-2D}
\begin{split}
    &\Psi(\vec{k_1},\vec{k_2})=\\
    &\sum_{mn}{\sqrt{\lambda_n\lambda_m}\psi_n(k_{1x})\psi_m(k_{1y}) \times \psi_n(k_{2x})\psi_m(k_{2y})}.
\end{split}
\end{equation}
Degree of entanglement for this two-dimensional wave-packet is given by Schmidt number:
\begin{equation}
    K=K_x\times K_y=(a^2+b^2)^2 / 4a^2b^2.
\end{equation}

Let us note, that (\ref{HG-Schmidt-2D}) is not the only possible
form of Schmidt decomposition. It may as well be described in
terms of Laguerre-Gaussian modes as in
\cite{EberlyPRL04,BramonPRA07}. The value of Schmidt number is, of
course, basis independent, as was explicitly shown in a recent
preprint by Miatto \emph{et al.} \cite{MiattoArXiv2011}. In fact,
the difference between these two representations corresponds to
the choice of polar or cartesian coordinates on the plane of
transverse momentum components.

\section{Self-calibrating tomography scheme for inferring Schmidt eigenvalues}
\label{self_calibrating}

Here we describe briefly the concept of self-calibration as
simultaneous updating of information about the state and the
measurement device as it was formulated in a recent work
\cite{mogilevtsev2012}. Also, we discuss the application of the
self-calibration scheme for tomography of angular Schmidt modes in
SPDC.

Generally, the possibility of self-calibration follows naturally
from the Born rule
\begin{equation}
p(\rho,X)=\mathrm{Tr}\{\hat{\Pi}(X)\rho\},
 \label{born rule}
\end{equation}
which gives one a probability of getting the particular measurement result, $p(\rho,X)$, being linear on coefficients of the representation of the
signal density matrix, $\rho$, and the elements of the Positive Operator Valued Measure (POVM),$\hat{\Pi}(X)$, in an arbitrary basis.
Obviously, \textit{a-priori} knowledge of some parts of the signal density matrix can be traded for getting knowledge of some
parts of the POVM. And even if one aims for inferring such parameters of the POVM (say, $X$ in Eq.(\ref{born rule})) that the probability,
$p(\rho,X)$, depends nonlinearly on, both $X$ and the signal density matrix, it can be estimated, provided that a measure of estimation success is
strictly convex with respect to all the parameters/coefficients to be found. For this measure one can take, for example, the Kullback-Leibler
divergence between the expected inferred set of probabilities, $\vec p=\{p_{ij}(\rho,X)\}$, and the actually measured frequencies, $\vec
f=\{f_{ij}\}$ \cite{mogilevtsev2012},
\begin{eqnarray}\label{loglik}
D(\vec p,\vec f)\propto \ln{(L)}=\sum\limits_{i,j}
{f_{ij}}\ln\left\{\frac{p_{ij}(\rho,X)}{P}\right\}, \\
\nonumber P=\sum_{ij} p_{ij}(\rho,X).
\end{eqnarray}
Minimizing this divergence is equivalent to performing
maximum-likelihood (ML) estimation \cite{mle_hradil}, i.e.
maximizing the likelihood $L$ of the model.

Building the self-calibration statistical estimation procedure for our case is greatly simplified, firstly, by the available \textit{a-priori}
knowledge about the signal state and, secondly, by the character of measurements. Actually, one needs estimating diagonal elements of the density
matrix (\ref{Schmidt-decomposition}) containing terms corresponding to no more than single photons. This allows one implementing highly efficient
iterative expectation-maximization algorithm for performing ML estimation of the diagonal elements of the density matrix, preserving positivity on
each step of the iteration procedure  \cite{em} (for practical implementation for the diagonal elements estimation see, for the example, Refs.
\cite{paris,mogilevtsev2009}). Also, as it will be seen below, the measurement is done by performing rather accurate (albeit still
significantly non-ideal) projection  on the chosen Schmidt mode approximated as the Gaussian function (see the previous Section). Local
measurements in one of directions are sufficient for the inference.

So, the reduced local signal density matrix describing one photon
of the pair is
\begin{equation}\label{single-rho}
    \rho(\vec{k},\vec{{k^\prime}}) = \sum_{n,m} \lambda_{mn} \psi_{nm}(\vec{k})\psi_{nm}(\vec{{k^\prime}}),
\end{equation}
where coefficients $\lambda_{nm}=\lambda_{m}\lambda_{n}$ are
Schmidt numbers corresponding to the mode $\psi_{nm}$. The POVM
elements for the local projective measurement of the mode
$\psi_{nm}$ can be written as
\begin{equation}
\Pi_{nm}(\vec{k},\vec{{k^\prime}})=\sum_{i,j} \mu^{(nm)}_{ij}
\psi_{ij}(\vec{k})\psi_{ij}(\vec{{k^\prime}}), \label{povm}
\end{equation}
where parameters $\mu_{ij}^{(nm)}$ describe losses and efficiency
of the projection. Knowledge about these parameters is to be
updated via the self-calibration procedure. The
expectation-maximization iterative procedure to the case can be
written as:
\begin{equation}
\lambda_{nm}^{(k+1)}=\lambda_{nm}^{(k)}\sum\limits_{i,j}\frac{f_{ij}}{p_{ij}^{(k)}}\frac{\mu^{(nm)}_{ij}}{\mu^{sum}_{ij}},
 \label{em}
\end{equation}
where
\[ \mu^{sum}_{ij}=\sum\limits_{n,m} \mu^{(nm)}_{ij},
\]
and probabilities $p_{ij}^{(k)}$ are estimated via the Born rule
(\ref{born rule}) on the $k$th iteration of the procedure. For the
initial approximation, $\lambda_{mn}^{(0)}$, one can choose, for
example, the set of equal numbers (really, the choice of
$\lambda_{mn}^{(0)}$ is not really important provided for that
they are non-zero \cite{paris}). An important feature of the
procedure (\ref{em}) specific to our case is that the result of
estimation should be factorable,
$\lambda_{nm}^{(k)}=\lambda_{m}^{(k)}\lambda_{n}^{(k)}$.
Practically, it can be done by replacing the result of each
iteration by the closest factorable matrix. As we shall see below,
the procedure (\ref{em}) provides for fast and efficient
estimation of Schmidt eigenvalues.

The self-calibration procedure can be imagined as follows: one
assumes some values of parameters $\mu_{ij}^{(nm)}$, estimates
$\lambda_{mn}$ and probabilities $p_{ij}$ and calculates the value
of the Kullback-Leibler divergence (\ref{loglik}). Then one
repeats the whole procedure for another set of $\mu_{ij}^{(nm)}$.
If the Kullback-Leibler divergence is convex in the chosen region
of parameters $\mu_{ij}^{(nm)}$, one chooses the values of
$\mu_{ij}^{(nm)}$ corresponding to the minimum divergence as ones
allowing the closest fitting of the experimental results.

\section{Experimental realization of measurements in Schmidt basis}
\label{experiment}

Here we describe locals measurement performed by projection to the
approximated Gaussian Schmidt modes.

We used a 2~mm BBO crystal pumped by a CW He-Cd laser with $\lambda_p=325$~nm wavelength. The crystal was cut for collinear frequency-degenerate
Type-I phase-matching. Angular bandwidth of phase-matching in such crystal (neglecting the pump divergence) -- $b$ parameter in
(\ref{double-gaussian}) is $\frac{\lambda}{2\pi}\times b=0.033$, where $\lambda=2\lambda_p=650$ nm is the wavelength of down-converted photons. It
was convenient for our purposes to select the value of pump divergence corresponding to a moderate Schmidt number. We focused the pump inside the
crystal with a 150~mm quartz lens and measured the divergence -- $a$ parameter in (\ref{double-gaussian}) to be $\frac{\lambda}{2\pi}\times
a=(5.8\pm0.1)\times10^{-3}$ \footnote{One should pay attention to the fact, that all formulas in previous sections do not take into account the
refraction on the crystal surface, so to compare with experiment all angular variables for the pump should be divided and for SPDC multiplied on
$n_e(\lambda_p)=n_o(\lambda_s)=1.667$.}. To ensure applicability of double-gaussian approximation we calculated eignevalues and eigenfunctions for
the reduced single-photon density matrix, corresponding to the precise SPDC wavefunction (\ref{Monken}) numerically. The calculation was performed
as follows: Hermite-Gaussian modes corresponding to the approximate function (\ref{double-gaussian-1D}) were chosen as a basis, we have restricted
ourselves to 10 lower order modes (giving the Schmidt number with 3 decimal digits precision) and calculated the matrix elements of the precise
density matrix in this basis. Diagonalizing the calculated matrix, we obtained eigenvalues and eigenfunctions. The results are in reasonable
correspondence with a simplified double-gaussian model, at least, we should expect that phase holograms for Schmidt modes should be close to those
of Hermite-Gaussian modes of appropriate divergence. We should note that measured waist size of the pump beam in the focal plane of
the lens was $w_p=(25\pm1)\,\mu \mathrm{m}$, corresponding to $M^2=1.4$. That means, the pump beam is aberrated and is not really gaussian, and that
may cause some deviations from Hermite-Gaussian shape of Schmidt modes as well.

\begin{figure}
\centering\includegraphics[width=\columnwidth]{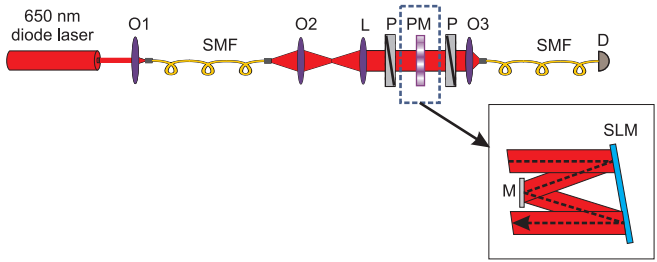}
\caption{{\protect{Experimental setup for transforming an
attenuated laser beam to higher Hermite-Gaussian modes. SMF -
single mode fiber; O1,2,3 - microscope objectives; L - 145 mm
lens; P - polarizers; D - single photon counter. A phase modulator
is shown as a transmitting mask, while in reality it is a
combination of SLM and a dielectric mirror M (see text for
details).}}} \label{Imitator}
\end{figure}

\begin{figure}
\centering\includegraphics[width=\columnwidth]{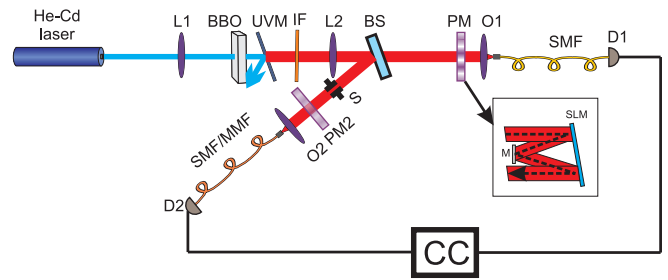}
\caption{{\protect{Experimental setup. L1 -- 150 mm quartz lens;
L2 -- 145 mm lens; BBO - 2 mm BBO crystal placed in the joint
focus of L1 and L2; UVM -- UV mirror cutting off the pump; IF --
interference filter for 650 nm with 40 nm bandwidth; BS --
non-polarizing 50/50 beam-splitter; 01,2 -- $8\times$ microscope
objectives; PM -- spatial light modulator (is shown as
transmitting mask for simplicity, real alignment is shown on the
inset); PM2 -- phase mask made of thin glass plates; SMF - single
mode fiber; SMF/MMF -- single or multi-mode fiber depending on the
experiment (see text for details); D1,2 -- single photon counters
(Perkin Elmer). A $200\, \mu m$ vertical slit S was used in
"ghost" imaging experiments.}}} \label{Setup-Shmidt}
\end{figure}

We have used an LCoS SLM with VAN matrix produced by Cambridge
Correlators. The matrix has $1027\times768$ pixels of $~10\,\mu
\mathrm{m}$ size each. It is an 8 bit device, capable of
introducing a phase shift of up to $0.8\pi$. Since larger phase
shifts are required for our holograms we used a double reflection
scheme. We used two polymer film polarizers in front and after the
SLM to reduce the unwanted polarization rotations by an additional
dielectric mirror necessary in such scheme (see insets in Fig.
\ref{Imitator},\ref{Setup-Shmidt}).

\begin{figure}
\centering\includegraphics[width=0.75\columnwidth]{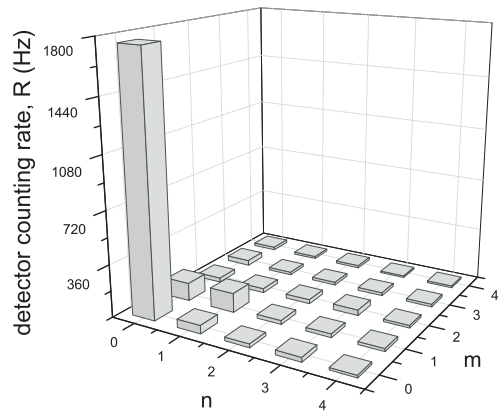}
\caption{{\protect{Detector counting rate for an attenuated laser
beam transformed to $HG_{nm}$ modes. The fiber tip is set to
position, corresponding to $x=0$ in
Fig.~\ref{Convolution-imitator}.}}} \label{Visibility-imitator}
\end{figure}

To estimate the quality of quality of mode transformation with this device we
used the setup sketched in Fig.~\ref{Imitator}.
We used an attenuated 650~nm diode laser as a source. The beam was mode filtered with single mode fiber and focused with a $20\times$ microscope
objective to obtain the divergence similar to that of an $HG_{00}$ Schmidt mode, and the waist at the position of the crystal. So we obtained a
single mode gaussian beam modeling the zero order Schmidt mode of SPDC beam. The beam was collimated with 145~mm lens and after reflection from SLM
was focused with a $8\times$ microscope objective to a single mode fiber followed by a single photon counter (Perkin Elmer). The focused beam waist
exactly coincided with the mode size of the fiber ($~4\, \mu \mathrm{m}$). We should stress, that we paid special attention to mode matching and
optics were chosen in such a way that detection mode exactly coincides with calculated $HG_{00}$ Schmidt mode. Parameters of phase holograms were
adjusted to minimize the detector counting rate, i.e. to ensure orthogonality of transformed modes to a fundamental gaussian one. We have actually
adjusted three parameters: the position of phase step for $HG_{10(01)}$ modes, which is determined by the beam position at the SLM (in horizontal
and vertical directions, respectively), and the distance between phase steps for $HG_{20}$ modes determined by beam size at the SLM plane. These
parameters define the shape of holograms for other modes in a unique way. If we define "visibility" for mode transformations as the ratio of
counting rates with holograms for $HG_{nm}$ modes to that for untransformed gaussian mode: $V=(R_{00}-R_{mn})/(R_{00}+R_{mn})$, then for almost all
of the modes with $0\leq m,n \leq 4$ it exceeds 97\%, corresponding to reasonably high quality of mode transformations. Histogram of counting rates
for various modes is shown in Fig.~\ref{Visibility-imitator}. Notice that counts rate for supposedly symmetrical modes $HG_{01}$ and $HG_{10}$ are
visibly different. It is a consequence of different quality of projections for these modes. Such non-symmetry should be accounted for by
correspondent POVM elements (namely, by the parameters $\mu_{ij}^{(nm)}$). This hardly controllable non-symmetry and other artifacts of non-perfect
mode transformations inevitable with phase-only holograms is the main reason for applying the self-calibrating reconstruction procedure for the
case.

\begin{figure}
\centering\includegraphics[width=\columnwidth]{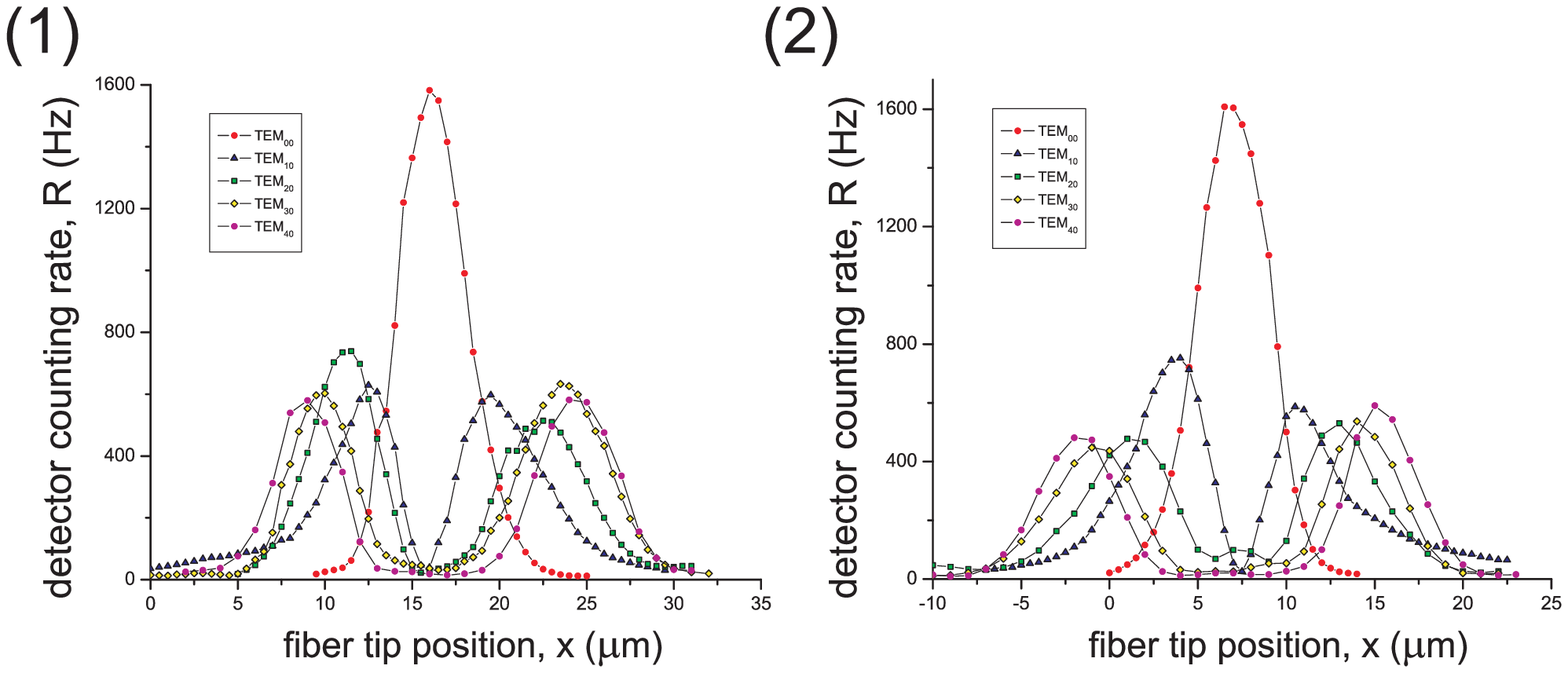}
\caption{{\protect{Counting rate dependence on the position of
fiber tip for an attenuated laser beam transformed to various
modes. Fiber is scanned in horizontal direction (1) and in
vertical one (2).}}} \label{Convolution-imitator}
\end{figure}

\begin{figure}
\centering\includegraphics[width=0.7\columnwidth]{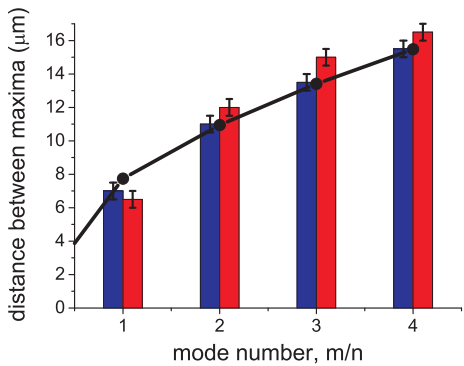}
\caption{{\protect{Dependence of maxima positions for fiber tip
scan on mode number for laser beam transformations. "Horizontal"
$HG_{n0}$ (blue bars), "vertical" $HG_{0m}$ (red bars) modes and
theoretical predictions (solid black line and dots). Theoretical
predictions are calculated for gaussian waist $a=3.9\, \mu
\mathrm{m}$ (see text for details).}}} \label{Maxima}
\end{figure}

To check whether the spatial structure of transformed modes is at least close to Hermite-Gaussian, we scanned the fiber tip in the focal plane of O3
objective. The counting rate dependence on fiber position is determined by the convolution of a corresponding Hermite-Gaussian function and a
fundamental gaussian mode of the fiber:
\begin{equation}\label{Convolution}
\begin{split}
    &R(x) \propto\\
    &\left|\int_{-\infty}^{\infty} \mathrm{H}_{nm}(\sqrt{2}\tilde{x}/w)\exp\left({-\frac{\tilde{x}^2}{w^2}}\right)\exp\left(-\frac{(x-\tilde{x})^2}{w^2}\right) d\tilde{x}\right|^2,
\end{split}
\end{equation}
where $w$ is the gaussian mode waist. Experimental dependencies
are shown in Fig.~\ref{Convolution-imitator} and have a
characteristic shape of double-peak curves. Distance between
maxima depends on the mode number, and is shown in
Fig.~\ref{Maxima} for "horizontal" $HG_{n0}$ and "vertical"
$HG_{0m}$ modes, together with theoretical predictions for
Hermite-Gaussian modes. To plot the theoretical predictions
correctly we estimated the waist size by fitting the convolution
for $HG_{00}$ mode with a gaussian curve, obtaining
$w=(3.87\pm0.07)\, \mu \mathrm{m}$.

When the attenuated laser beam is substituted with SPDC radiation, the described scheme realizes projective measurements in approximately
Hermite-Gaussian basis. Full scheme of experimental setup is shown in Fig.~\ref{Setup-Shmidt}. Pump was focused to a 2 mm BBO crystal with a 150 mm
quartz lens L1, a second lens L2 with $F=145$~mm was set confocal with L1 to collimate the beam. Pump radiation was cut off with a UV-mirror UVM,
and SPDC radiation was frequency filtered with an interference filter IF. We used filters with central wavelength of 650 nm and bandwidth of 40 nm
and 10 nm and did not observe any significant improvement of visibility for narrower filter. All the following results were obtained with a wide 40
nm filter. Photon pairs were split with a 50/50 non-polarizing beam-splitter. An SLM was placed in the transmitted channel and after reflection the
radiation was focused into single mode fiber placed in the focal plane of $8\times$ microscope objective. In the reflected channel the beam was
focused into similar single mode fiber with identical objective. Signals in both channels were detected by single photon counters connected to a
coincidence circuit.

\begin{figure}
\centering\includegraphics[width=\columnwidth]{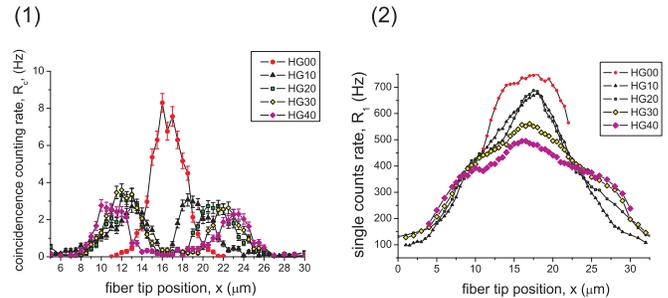}
\caption{{\protect{Coincidence (1) and single counts (2) rate
dependence on the fiber tip position in the focal plane of the
microscope objective in the channel with SLM for different modes.
Fiber is scanned in horizontal direction.}}}
\label{Convolution-SPDC}
\end{figure}

\begin{figure}
\centering\includegraphics[width=0.75\columnwidth]{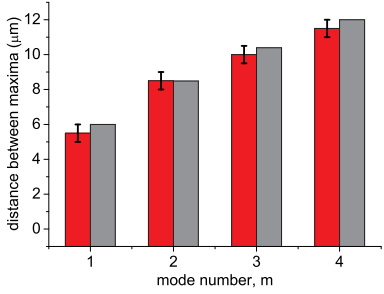}
\caption{{\protect{Dependence of maxima positions for coincidence
distributions of Fig.~\ref{Convolution-SPDC} on the mode number
(red bars). Calculated dependence for Hermite-Gaussian modes with
$w=(3.0\pm0.1)\, \mu m$ corresponding to $HG_{00}$ waist size
(grey bars) is provided for comparison.}}} \label{Maxima-SPDC}
\end{figure}

Experimental evidence of similarity of Schmidt modes to
Hermite-Gaussian ones may be obtained by analyzing the
dependencies of single counts and coincidences on the fiber tip
position in the focal plane of the focusing microscope objective.
We expect the dependence for coincidences to be described by
(\ref{Convolution}). Experimental curves for the case when fiber
in the transmitted channel is scanned are shown in
Fig.~\ref{Convolution-SPDC}. Distance between maxima behaves analogously to
the case of attenuated laser beam, as shown in
Fig.~\ref{Maxima-SPDC}.

\begin{figure}
\centering\includegraphics[width=0.75\columnwidth]{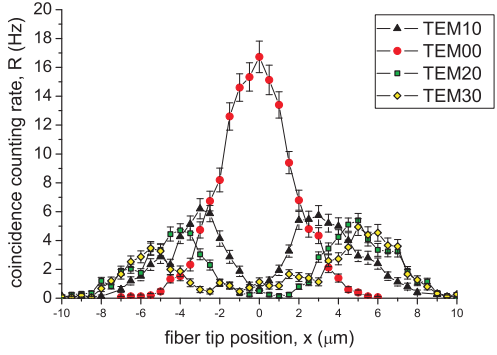}
\caption{{\protect{Coincidence counting rate dependence on the
fiber tip position in the reflected channel (without SLM) for
different modes. Fiber tip is scanned in horizontal direction.}}}
\label{reflected-scan}
\end{figure}

We obtained same dependencies of coincidence counting rate when
the fiber tip was scanned in the reflected channel (see
Fig.~\ref{reflected-scan}). In this case single counts, obviously,
do not depend on the mode selected in the conjugate channel at
all. This effect is a straightforward consequence of intermodal
correlations in SPDC and may be thought of as a sort of "ghost
interference" \cite{KulikPRL04}. We should note that almost zero
coincidence counting rate in the central position of the fiber is
an interference effect demonstrating spatial coherence of detected
modes. So this result may be considered as an experimental
demonstration of one of the main features of Schmidt modes --
their spatial coherence.

Note that the double-peak structure characteristic for Hermite-Gaussian modes appears only in coincidences dependence, while single counts behave
monotonously, as is expected for spatially multi-mode radiation. The maximal value of single counting rate, however, decreases with increasing value
of mode indexes.This is clear from the form of single-photon density matrix, described by (\ref{single-rho}). With the fiber placed in central
position, the detection scheme in the transmitted arm of the setup realizes projections described by (\ref{povm}), single counts rates in this case
correspond to frequencies $f_{ij}$ in (\ref{em}) and are the data used for statistical inference of Schmidt eigenvalues.

{\section{Performing self-calibration for inference of Schmidt
eigenvalues}}

Essential features of the measurement procedure described in the
previous Section can be captured by writing down parameters of
POVM (\ref{povm}) in the following form
\begin{equation}
\mu_{ij}^{(nm)}\approx\mu_{tot}\eta_{nm}\Gamma_{ij}^{(nm)}.
\label{povm particular}
\end{equation}
In Eq.(\ref{povm particular}) the parameter $\mu_{tot}$ represent total losses (detection efficiency) equal for all measured modes. Since it
contributes to normalization of the estimated signal density matrix only, it is irrelevant. The matrix $\eta_{nm}$ represents asymmetry of losses.
Thus, we assume $\eta_{nm}\equiv1$ for $n\leq m$. The parameters $\eta_{nm}$ for $n\geq m$ are to be determined via the self-calibration procedure.
Parameters $\Gamma_{ij}^{(nm)}$ describe quality of the projection  Since it was demonstrated that the projection is of a
reasonably good quality, we assume
\begin{equation}
\Gamma_{ij}^{(nm)}\approx
\Delta_{mn}\delta_{in}\delta_{jm}+d_{ij}^{(nm)}, \label{projection
quality}
\end{equation}
where all the parameters in the right-hand side of formula
(\ref{projection quality}) are taken to be non-negative, and for
each $m$, $n$ the quantity
\begin{equation}
 \Delta_{mn}\gg \sum\limits_{i,j}d_{ij}^{(nm)}.
 \label{delta_condition}
\end{equation}
Also, for simplicity sake we assume that inefficiency of the
projection occurs solely from contribution of other modes, so
$\sum\limits_{i,j}\Gamma_{ij}^{(nm)}=1$. Parameters $\Delta_{mn}$
can be chosen from results of projection quality measurements
described in the previous Section. Thus, we have taken
$\Delta_{mn}=0.97$ for all $m,n$ apart from
$\Delta_{01}=\Delta_{11}=0.9$. For simulations small parameters
$d_{ij}^{(nm)}$ were sampled randomly from the homogeneous
distribution.

\begin{figure}
\centering\includegraphics[width=1.2\columnwidth]{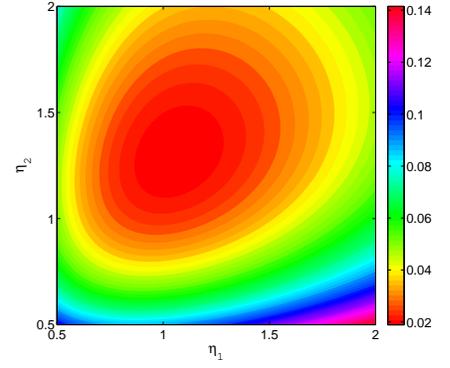}
\caption{{\protect{The Kullback-Leibler divergence
(\ref{kullback}) for the experimentally obtained frequencies.}}}
\label{dmfig1}
\end{figure}

\begin{figure}
\centering\includegraphics[width=1.3\columnwidth]{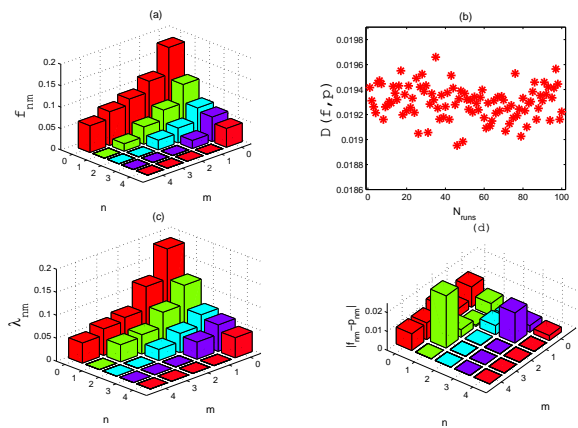}
\caption{{\protect{The panel (a) shows the registered frequencies
of photocounts for different modes $HG_{nm}$. The panel (b) shows
variations of the estimated minimal Kullback-Leibler divergence
for different realizations of the randomly chosen parameters
$d_{ij}^{(nm)}$; $N_{runs}$ denoted the number of the particular
realization. The panel (c) shows the the reconstructed signal
density matrix elements $\lambda_{nm}$ for asymmetry parameters,
$\eta_1=1.3$ and $\eta_2=1.125$. In the panel (d) the absolute
value of differences between experimentally measured frequencies
and the estimated probabilities is shown for values of parameters
as for the panel (c).}}} \label{dmfig2}
\end{figure}

To model asymmetry of losses, we have introduced two parameters,
$\eta_1$ and $\eta_2$. We assume that elements of the matrix
$\eta_{nm}$ closest to the main diagonal are equal to $\eta_1$.
Other elements of $\eta_{nm}$ for $n\geq m$ are taken to be equal
to $\eta_2$. In Fig.\ref{dmfig1} the Kullback-Leibler divergence
(\ref{loglik}) is shown for the different values of parameters
$\eta_{1,2}$; for calculations Eq.(\ref{loglik}) is recast in the
standard form
\begin{eqnarray}
D(\vec p,\vec f)=\sum\limits_{i,j}
{f_{ij}}\ln\left\{\frac{f_{ij}}{p_{ij}(\rho,X)}\right\},
\label{kullback}
\end{eqnarray}
where both sets of estimated probabilities $\vec p$ and measured
frequencies  $\vec f$ are normalized to unity. The calculation is
done for experimentally obtained  set of frequencies shown in
Fig.\ref{dmfig2}(a) using the iterating ML estimaion procedure
described in the Section \ref{self_calibrating} and given by
Eq.(\ref{em}). One can see that the Kullback-Leibler divergence is
obviously convex for the chosen range of parameters $\eta_{1,2}$.
The minimum is reached for $\eta_1\approx 1.3$,
$\eta_2\approx1.125$.

The result of the signal state estimation with the inferred
parameters of the detecting scheme is given in
Fig.\ref{dmfig2}(c). It is symmetric, i.e.
$\lambda_{nm}=\lambda_{mn}$ (and it is rather different from the
experimentally found frequencies shown in Fig.\ref{dmfig2}(a)).
However, the reconstructed set of $\lambda_{nm}$ gives a set of
probabilities rather close to the measured frequencies
(Fig.\ref{dmfig2}(d)). The reconstruction procedure is robust with
respect to small imperfections in performing projections. It is
seen (see Fig.\ref{dmfig2}(b)) that as long as the condition
(\ref{delta_condition}) holds, the result of the estimation
changes rather weakly for different realizations of random
variables  $d_{ij}^{(nm)}$.

Thus, we have established that the self-calibration procedure
allows one to infer with rather high precision the quantum state
entering the detection scheme and parameters of this detection
scheme. Now let us consider how the inferred state agrees with the
double-Gaussian model developed in the Section \ref{SPDC_state}.

\begin{figure}
\centering\includegraphics[width=1.1\columnwidth]{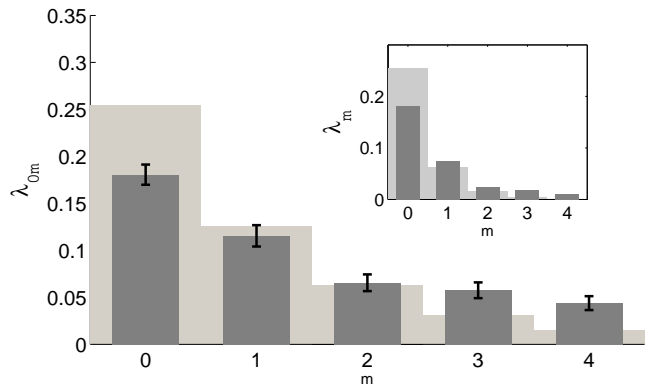} \caption{{\protect{An estimated edge elements $\lambda_{0m}$(dark gray bars) and
the ones modelled according to Eq.(\ref{HG-Schmidt-eigenvalues}) (light gray bars). Errors bars represent the deviation estimated using Fisher
matrix method (\ref{varfisher}). The inset shows the estimated (dark gray bars) and the modelled (light gray bars) Schmidt eigenvalues corresponding
to the parameters $\frac{\lambda}{2\pi}\times a=5.8\times10^{-3}$, $\frac{\lambda}{2\pi}\times b=0.0334$.}}} \label{dmfig3}
\end{figure}

In Fig.\ref{dmfig3} one can compare estimated values of
$\lambda_{0m}$ and ones obtained using the double-Gaussian model
Eq.(\ref{HG-Schmidt-eigenvalues}) with parameters $a$ and $b$
determined in the previous Section. It can be seen that while
some modeled and estimated eigenvalues are rather close,
difference between other ones goes beyond estimation errors
determined for MLE procedure via the Fisher information matrix
\cite{mognjp2008}. The variance of the inferred element
$\lambda_{nm}$ can be estimated as
\begin{eqnarray}
\Delta_{nm}=F^{-1}_{nm,nm},
 \label{varfisher}
\end{eqnarray}
where the Fisher information matrix is
\begin{eqnarray}
\nonumber
F_{kl,mn}=\sum\limits_{j=0}^N\frac{N_{mes}}{f_{j}}\frac{\partial}{\partial
\lambda_{kl}}\left[{p_{j}}\right]\frac{\partial}{\partial
\lambda_{mn}}\left[{p_{j}}\right],
\end{eqnarray}
$N_{mes}$ being the total number of measurements ($1,800$ measurements per mode was actually taken in the experiment). The
estimated value of the Schmidt number, $K_{x,y}=3.34$, is reasonably close to the modelled value, $K_{x,y}^{est}\approx 2.97$.

It seems that the discrepancy can be explained, firstly, by
systematic fluctuations of signal level on long time scales caused
primarily by temperature fluctuations. Experiments have shown,
that both coincidence and single counts rates are extremely
sensitive to small beam displacements. So work of an
air-conditioning system stabilizing the room temperature caused
small but noticeable periodic change of signal with a period of
$~10$ min.  Complete thermal isolation of the setup should remove
this source of errors. Another significant source of errors may be
the difference between detected angular aperture and full angular
bandwidth of SPDC. Sharp dependence of counting rates on the
position of fiber tip relatively to the microscope objective shows
maximum in the position slightly different from the focal plane.
Moreover, the pump, assumed to be Gaussian, was actually aberrated
to $M^2=1.4$, which may also slightly change the real
distribution.

Nevertheless, we believe our results to be providing sufficiently persuasive evidence of possibility to realize projective measurements in a basis
close to spatial Schmidt modes basis for SPDC biphotons, and to perform the self-calibrating procedure for estimating both the parameters of the
measurement set-up and the signal SPDC state.

\section{Conclusion}

In this work we have presented the first experimentally realized self-calibrating procedure for simultaneous estimation of the signal SPDC state and
imperfections of the detecting set-up. To this end we have analyzed spatial entanglement in SPDC in terms of spatial Schmidt decomposition and shown
that under reasonable assumptions, applicable to the particular experiment, these modes are close to Hermite-Gaussian modes. Using this fact we have
experimentally realized a scheme of projective measurements in approximately Hermite-Gaussian basis using an active spatial light modulator.
Experimental results prove high quality of Gaussian beam transformation to higher order Hermite-Gaussian modes. For a spatial multi-mode SPDC
radiation such spatial filtering allowed us to realize projective measurements approximating projections in Schmidt basis.

We have developed the self-calibrating protocol for the case and
have demonstrated that the predicted probabilities are quite close
to the registered frequencies. It clearly demonstrates validity
of the self-calibration procedure performed for the case. Despite
some discrepancies between the predictions of the double-Gaussian
model for the Schmidt modes and the the results of the inference
on the basis of the experimentally measured data, one can also
claim that the double-Gaussian model is a good choice for
modeling actual Schmidt modes of the generated SPDC state. Thus,
we conclude that the self-calibrating procedure can become a
powerful tool for updating information about the measurement
itself while performing the diagnostics of the generated
non-classical state.

We are grateful to M.~V.~Fedorov for stimulating discussions. This work was supported in part by Federal Program of the Russian Ministry of
Education and Science (grant 8393), European Union Seventh Framework Programme under grant agreement n° 308803 (project
BRISQ2), ERA.Net RUS Project “NANOQUINT”, grant for the NATO project EAP.SFPP 984397 "Secure Communication Using Quantum Information Systems" and by RFBR grant 12-02-31041. S.~S.~Straupe and I~.B.~Bobrov are grateful to the "Dynasty" foundation for financial support. This work was also supported by the Foundation of Basic Research of the Republic of Belarus, by the National Academy of Sciences of Belarus through the Program "Convergence", by the Brazilian Agency FAPESP (project 2011/19696-0) (D.M.)

\label{conclusion}

\end{document}